# Accelerated Coronary MRI with sRAKI:

# A Database-Free Self-Consistent Neural Network

# k-space Reconstruction for Arbitrary Undersampling

Seyed Amir Hossein Hosseini[1,2], Chi Zhang[1,2], Sebastian Weingärtner[1,2,3],

Steen Moeller[2], Matthias Stuber[4,5], Kâmil Uğurbil[2], and Mehmet Akçakaya[1,2]

[1]Electrical and Computer Engineering, University of Minnesota, Minneapolis, MN
[2]Center for Magnetic Resonance Research, University of Minnesota, Minneapolis, MN
[3]Department of Imaging Physics, Delft University of Technology, Delft, Netherlands
[4]Department of Radiology, University Hospital (CHUV) and University of Lausanne (UNIL), Lausanne, Switzerland
[5]Center for Biomedical Imaging (CIBM), Lausanne, Switzerland

**Funding:**

NIH, Grant numbers: R00HL111410, P41EB015894, U01EB025144, P41EB027061; NSF, Grant number: CAREER CCF-1651825

# Abstract

**Purpose:** To accelerate coronary MRI acquisitions with arbitrary undersampling patterns by using a novel reconstruction algorithm that applies coil self-consistency using subject-specific neural networks.

**Methods:** Self-consistent robust artificial-neural-networks for k-space interpolation (sRAKI) performs iterative parallel imaging reconstruction by enforcing self-consistency among coils. The approach bears similarity to SPIRiT, but extends the linear convolutions in SPIRiT to nonlinear interpolation using convolutional neural networks (CNNs). These CNNs are trained individually for each scan using the scan-specific autocalibrating signal (ACS) data. Reconstruction is performed by imposing the learned self-consistency and data-consistency, which enables sRAKI to support random undersampling patterns. Fully-sampled targeted right coronary artery MRI was acquired in six healthy subjects. The data were retrospectively undersampled, and reconstructed using SPIRiT, $\ell_1$-SPIRiT and sRAKI for acceleration rates of 2 to 5. Additionally, prospectively undersampled whole-heart coronary MRI was acquired to further evaluate reconstruction performance.

**Results:** sRAKI reduces noise amplification and blurring artifacts compared with SPIRiT and $\ell_1$-SPIRiT, especially at high acceleration rates in targeted coronary MRI. Quantitative analysis shows that sRAKI outperforms these techniques in terms of normalized mean-squared-error (~44% and ~21% over SPIRiT and $\ell_1$-SPIRiT at rate 5) and vessel sharpness (~10% and ~20% over SPIRiT and $\ell_1$-SPIRiT at rate 5). Whole-heart data shows the sharpest coronary arteries when resolved using sRAKI, with 11% and 15% improvement in vessel sharpness over SPIRiT and $\ell_1$-SPIRiT, respectively.

**Conclusion:** sRAKI is a database-free neural network-based reconstruction technique that may further accelerate coronary MRI with arbitrary undersampling patterns, while improving noise resilience over linear parallel imaging and image sharpness over $l_1$ regularization techniques.

# Introduction

Coronary artery disease (CAD) is the leading cause of death in the United States, accounting for one in seven deaths[1]. Coronary MRI provides a non-invasive and radiation-free diagnostic tool for CAD assessment[2], with a potential for repeated use. It is typically acquired with electrocardiogram (ECG) triggering during diastolic quiescence, where ~30-35 k-space lines are sampled per R-R interval[3–5]. When imaging the right coronary artery in a targeted manner[3], this leads to a ~3 minute nominal scan time. Since this scan time necessitates a free-breathing acquisition[6,7], respiratory motion compensation needs to be applied[4,5], typically with navigator gating[5,8], which further reduces the efficiency of the scans by ~2-3 fold, leading to a scan time of ~6-10 minutes. Alternatively, coronary MRI can be acquired with whole-heart coverage, which leads to a higher signal-to-noise ratio (SNR)[9,10], albeit at a longer nominal acquisition time of 6-8 minutes. The additional scan time overhead due to respiratory motion compensation often requires accelerated acquisitions, necessitating a trade-off with SNR[9,11].

Several strategies have been used to accelerate coronary MRI acquisitions such as parallel imaging[12,13], compressed sensing[14–16], and their combinations[17–23]. Recently, deep learning-based techniques[24–35] have also gained attention as a means to accelerate MRI acquisition. Numerous studies have designed neural network architectures that either establish an end-to-end nonlinear mapping from under-sampled k-space/distorted image to full k-space/undistorted image[25,27,28,31,33–35] or decompose an iterative optimization problem into (recurrent) deep learning blocks that learn a data-specific regularization[26,29,30,32]. A number of these studies also show support for parallel imaging with multi-coil data[24,26,29,31]. While these studies show promising results in accelerated MRI, there are limitations regarding the training phase of reconstruction. First, large datasets are required for training the neural networks which is not readily available in all situations. Second, it is infeasible to acquire fully-sampled training data in some applications, e.g., in whole-heart coronary MRI, where the scan time would become prohibitively long. Furthermore, training datasets may not include all pathologies of interest, which may lead to risks in generalizability for diagnosis[36]. These obstacles may hinder the clinical application of current deep learning-based scan time acceleration techniques to high-resolution cardiac MRI[36].

An alternative line of work considers subject-specific application of neural networks[24]. In this approach, called robust artificial-neural-networks for k-space interpolation (RAKI), several convolutional neural networks (CNN) are calibrated from scan-specific autocalibrating signal (ACS) data for improved interpolation of missing k-space lines. Thus, this method extends the linear convolutions used in GRAPPA[37], and was shown to increase noise resilience for uniform undersampling patterns, especially in low-SNR and high-acceleration rate regimes[24]. However, previous work has shown the benefits of random undersampling in high-resolution three-dimensional (3D) coronary MRI, for instance in the setting of compressed sensing[17]. For such undersampling patterns, iterative self-consistent parallel imaging reconstruction (SPIRiT)[38] provides a k-space interpolation approach for multi-coil data. SPIRiT utilizes multi-coil information by including a self-consistency term that ensures the interpolated k-space is consistent with itself according to the calibration kernels, along with a data-consistency term in reconstruction. SPIRiT requires iterative processing in the reconstruction and is consequently more computationally-intensive than GRAPPA.

In this study, we exploit the notion of coil self-consistency in SPIRiT to enable RAKI with arbitrary undersampling. The proposed technique, called self-consistent RAKI (sRAKI), is evaluated in targeted and whole-heart coronary MRI, and compared with SPIRiT and $\ell_1$-SPIRiT at various acceleration rates. This work has been partially presented in[39–42].

# Methods

## Calibration

For multi-coil k-space data with $n_c$ coils, a k-space point in the $j^{\text{th}}$ coil, $x_j(k_x, k_y, k_z)$ can be estimated as a function of distinct k-space points from all coils $i \in \{1, ..., n_c\}$ within a neighborhood region of $(k_x, k_y, k_z)$[37,38]. In linear parallel imaging techniques, this function is modeled by a linear spatially shift-invariant convolution, and the convolutional kernels can be found by solving $n_c$ linear least squares optimization problems[37,38]. In particular, SPIRiT uses these linear convolutional kernels to define a coil self-consistency rule that connects all the k-space elements with neighboring elements across all coils. However, it has been noted that a nonlinear mapping may be advantageous from a noise reduction perspective due to two factors[24,43]. First, the shape and size of the neighborhood is heuristically set in practice[24], which may not capture all the required dependencies. Second, in contrast to typical least squares optimization problems, both the target and source points for the kernels in calibration are contaminated with noise[24,43], and nonlinear functions have been shown to deal more effectively with such imperfections[24,43]. Thus, we propose to utilize CNNs that are calibrated on ACS data of a single scan only to nonlinearly model the self-consistency in multi-coil k-space data.

In this study, a 4-layer CNN architecture was employed to learn the self-consistency rule among coils (**Fig. 1**). In contrast to conventional RAKI, where separate CNNs were used for mapping to individual coils, a single CNN was used to map from all coils of multi-coil k-space onto itself, facilitating considerably reduced run time. For reduced computational complexity, 3D k-space data was first inverse Fourier transformed along fully-sampled $k_x$ dimension[44,45]. Subsequently 2D convolutional kernels were jointly calibrated on the resultant 2D slices of data[44]. The k-space data across all coils were normalized to have unit power as a preprocessing step to enable the use of a fixed learning rate. In addition, the complex k-space data was embedded to the real field, by concatenating the real and imaginary components of k-space along the coil dimension leading to $2n_c$ input and output channels. All layers, except the last one, were followed by rectifier linear units (ReLU) as activation functions. The kernel size at input and output layers was 5×5, while the hidden layers used 3×3 kernels. The number of output channels of different layers was 16, 8, 16 and $2n_c$, respectively. The network was trained by minimizing a MSE objective function using Adam optimizer[46]. A learning rate of 0.01 and maximum number of iterations of 1000 were used in training.

**Reconstruction**

After calibrating the CNN on ACS data to learn the coil self-consistency rule, the following objective function is minimized to reconstruct k-space:

$$\arg\min_{\mathbf{x}} \|\mathbf{y} - \mathbf{D}\mathbf{x}\|_2^2 + \beta \|\mathbf{x} - \mathbf{G}(\mathbf{x})\|_2^2, \qquad (1)$$

where $\mathbf{x}$ is the reconstructed k-space data across all coils, $\mathbf{y}$ is the noisy acquired data, $\mathbf{D}$ is the undersampling operator and $\mathbf{G}(\cdot)$ represents the CNN for self-consistency. The first term in the objective function in (1) ensures that the reconstructed k-space is consistent with acquired data. The second term enforces self-consistency in the reconstructed k-space according to the coil self-consistency rule that was learned by calibrating on the ACS data. The parameter $\beta$ determines the balance between these two terms. Note that the main difference between sRAKI and RAKI is in this phase, where RAKI performs a one-time application of calibration kernels to estimate the missing data, whereas sRAKI requires iterative optimization of Equation (1). Additional regularization terms can also be incorporated in (1), although this was not investigated in the current study to maintain the focus on multi-coil data processing.

The objective function in (1) was optimized using the Adam optimizer with a heuristically chosen learning rate of 2, for the same k-space normalization to unit power as before. During this phase, the gradients need to be calculated with respect to **x,** i.e. the network input rather than network parameters, which is efficiently done through back-propagation. In order to avoid a heuristic tuning of $\beta$, consistency with data was strictly enforced as in SPIRiT[38]. This led to gradients being calculated for non-acquired elements only while the rest of k-space was directly replaced with acquired data at each iteration. For comparison, SPIRiT using a conjugate gradient reconstruction was implemented with a 5×5 kernel[38]. $\ell_1$-SPIRiT was also implemented with additional a Daubechies-wavelet regularization[38], where the thresholding parameter was empirically tuned to 0.0005 of the maximum absolute wavelet coefficient. The number of reconstruction iterations were tuned separately for each technique and was set to 50 for SPIRiT and sRAKI, and 15 for $\ell_1$-SPIRiT due to faster convergence of the latter.

## Targeted Coronary MRI

All imaging was performed on a 3T Siemens Magnetom Prisma (Siemens Healthineers, Erlangen, Germany) system with a 30-channel receiver body coil-array. The imaging protocols were approved by the local institutional review board, and written informed consent was obtained from all participants before each examination for this HIPAA-compliant study.

Targeted right coronary artery (RCA) MRI was acquired on 6 healthy subjects (26.7 ± 2.9 years, 3 women). Scout images were followed by axial breath-hold cine bSSFP images to identify the quiescent period of the RCA, which was used for the trigger delay of coronary acquisitions. A low-resolution free-breathing ECG-triggered 3D coronary survey was acquired for slab orientation of the RCA imaging. Targeted RCA MRI was then acquired with a free-breathing ECG-triggered GRE sequence with imaging parameters: TR/TE=3.4/1.5ms, flip angle=20°, bandwidth=601 Hz/pixel, field-of-view (FOV)=300×300×48 mm³, resolution=1×1×3 mm³, navigator window=5 mm. The nominal scan time was 160 seconds at a heart rate of 60 bpm. $T_2$-preparation and a spectrally-selective fat saturation were utilized for improved contrast.

The 3D k-space data was exported and retrospectively undersampled with a Poisson disc pattern at acceleration rates 2, 3, 4, and 5 with a fully-sampled $40 \times 10$ ACS region in $k_y - k_z$ plane. These under-sampled data were then reconstructed using SPIRiT, $\ell_1$-SPIRiT and sRAKI for comparison, with the implementations detailed above. Final images were obtained using root-sum-squares combination of all coil images. All algorithms were implemented in Python, and processed on a workstation with an Intel E5-2640V3 CPU (2.6GHz and 256GB memory), and an NVIDIA Tesla V100 GPU with 32GB memory.

**Image Analysis**

Quantitative analysis of the reconstructions was performed using normalized mean square error (NMSE) with respect to the fully-sampled reference, as well as normalized vessel sharpness measurements. NMSE was calculated in image domain between a given reconstruction method and the fully-sampled reference, normalized by the energy of the reference. Vessel sharpness scores were calculated for both sides of the vessel using a Deriche algorithm[47]. Normalized vessel sharpness was calculated as the average score of both sides divided by the intensity at vessel center. A normalized vessel sharpness value closer to 1 represents a sharper vessel border. The NMSE and normalized vessel sharpness measurements of the different reconstructions were statistically compared using paired t-test for each acceleration rate. A *p*-value of <0.05 was considered significant.

**Whole-Heart Coronary MRI**

Prospectively undersampled whole-heart coronary MRI was acquired on an additional subject (28 years, male) at an acceleration rate of 5 with a Poisson disc pattern. The same sequence parameters were used with FOV=300×300×106 mm$^3$, resolution=1.2×1.2×1.2 mm$^3$. The data were then reconstructed using SPIRiT, $\ell_1$-SPIRiT and sRAKI for comparison, with the same implementations described above. We note that this scenario poses a challenge for traditional machine learning algorithms that perform supervised learning on databases of fully-sampled data, as it is difficult to acquire high-quality fully-sampled whole-heart coronary MRI data. This is due to the long scan time of a fully-sampled acquisition, which leads to quality degradation due to drift and changes in the motion patterns. We also note that there have been some recent efforts to acquire fully-sampled whole-heart coronary MRI for this purpose, even though the acquisition time remains long[48].

# Results

**Fig. 2** depicts reformatted images from a targeted coronary MRI dataset reconstructed using SPIRiT, $\ell_1$-SPIRiT and sRAKI techniques at retrospective acceleration rates 2, 3, 4, and 5. RCA is visualized at all rates for all methods. sRAKI has visibly less noise at high acceleration rates compared to SPIRiT and fewer blurring artifacts compared to $\ell_1$-SPIRiT. The reformatted images from a second subject, are shown in **Fig. 3** with similar results showing that sRAKI has visibly less noise at high acceleration rates. sRAKI demonstrates improved quality at higher acceleration rates, reducing noise amplification and blurring artifacts compared with other reconstruction methods.

**Fig. 4** summarizes the mean and standard deviation of the NMSE and normalized vessel sharpness measurements for SPIRiT, $\ell_1$-SPIRiT and sRAKI across all subjects. sRAKI improves mean NMSE by 34%, 30%, 39%, 44% compared to SPIRiT, and 18%, 21%, 21% and 21% compared to $\ell_1$-SPIRiT for rates 2, 3, 4 and 5, respectively. Statistical analysis confirms that sRAKI significantly improves NMSE at all acceleration rates over SPIRiT (*p*-values: 0.002, 0.001, 0.004, 0.008, for given rates) and also significantly improves NMSE over $\ell_1$-SPIRiT at rates 4 and 5 (*p*-values: 0.125, 0.061, 0.036 and 0.013, for given rates). In terms of normalized vessel sharpness, sRAKI provides 7%, 9%, 11%, 10% improvement compared to SPIRiT and 4%, 5%, 13% and 20% improvement compared to $\ell_1$-SPIRiT for rates 2 to 5, respectively. The improvements over SPIRiT are statistically significant at rates 2 and 4 (*p*-values: 0.032, 0.115, 0.047 and 0.094, for given rates), while improvements over $\ell_1$-SPIRiT are statistically significant at rates 4 and 5 (*p*-values: 0.302, 0.139, 0.026 and 0.002, for given rates).

**Fig. 5** depicts the results of a prospectively 5-fold accelerated whole-heart coronary imaging. sRAKI yields a higher contrast and a sharper visualization of both the RCA and the left circumflex artery (LCX) compared to SPIRiT and $\ell_1$-SPIRiT. The normalized vessel sharpness measurements for this subject were 0.30, 0.31 and 0.33 for RCA and 0.25, 0.22, 0.28 for LCX with SPIRiT, $\ell_1$-SPIRiT and sRAKI reconstructions.

# Discussion

In this study, we proposed a novel reconstruction method called sRAKI to accelerate coronary MRI. sRAKI trained subject-specific CNNs to learn a nonlinear coil self-consistency rule for multi-coil k-space data. In the reconstruction phase, this learned self-consistency rule was enforced along with data-consistency constraints, similar to SPIRiT reconstruction. Thus, sRAKI enabled reconstruction with arbitrary undersampling patterns, an extension to RAKI[24], which was designed to handle uniform undersampling patterns only. In contrast to the recent machine learning-based MRI techniques[25–35], which require large training datasets, sRAKI is trained on subject-specific ACS data. This feature is especially advantageous in applications where fully-sampled training data cannot be acquired due to impractically long scan durations, such as whole-heart coronary imaging[9–11]. In addition, training CNNs on subject-specific data ensures inherent inter-subject variabilities of data are fully considered in the training process[36].

Several modifications were made to RAKI[24]. First, RAKI employed separate CNNs to learn nonlinear mapping functions from zero-filled multi-coil k-space data to missing data of individual coils. Therefore, $2n_c$ CNNs were trained to learn a full mapping function from multi-coil data to itself. In the new setting, we exploited a single CNN with more hidden layers to learn the coil self-consistency rule jointly, considerably reducing run time. Second, RAKI was examined in only 2D scenarios, whereas sRAKI was implemented for 3D datasets with two phase encoding dimensions. Another major difference is concerned with the reconstruction phase in which RAKI interpolates missing data with no iterations, but sRAKI optimizes an objective function to enforce data-consistency and self-consistency among coils. This procedure, which is similar to the reconstruction phase of SPIRiT, increases the computational burden by requiring first-order derivative calculation in each iteration. However, the extra complexity is not limiting. In this study, calibration on targeted right coronary artery datasets took ~20 seconds for SPIRiT and $\ell_1$-SPIRiT, and ~40 seconds for sRAKI all on GPU implementations, although none of the implementations were fully optimized. In addition, the reconstruction phase on GPU took ~220, 120 and 100 seconds for SPIRiT, $\ell_1$-SPIRiT and sRAKI, respectively.

In this study, the CNN parameters including the number of layers, the number of layer output channels and kernel sizes were empirically set. While our results indicate that a fixed parameter set yields satisfactory results in all datasets with various undersampling patterns, further optimization may improve the reconstruction.

Similar to SPIRiT, regularization terms can be included in the sRAKI objective function, in order to incorporate additional prior information, such as sparsity in transform domains[14–16]. However, these regularization parameters often need to be carefully tuned to avoid residual artifacts[16]. On the other hand, sRAKI without transform domain regularization, whose objective function requires no additional parameter tuning, showed desirable noise properties. The noise improvement in sRAKI is learned from the coil geometry, and does not inherently include any assumptions about compressibility in transform domains. A combination of sRAKI with advanced regularizers bears potential for improved reconstruction quality in certain scenarios, but was beyond the scope of this work, which emphasized the multi-coil aspect of the data.

# Conclusion

The proposed sRAKI reconstruction is a database-free CNN-based technique for self-consistent parallel imaging with arbitrary undersampling patterns, where the CNNs are trained on scan-specific ACS data. sRAKI is effective in accelerating coronary MRI, and improves reconstruction quality compared to regularized and non-regularzied SPIRiT.

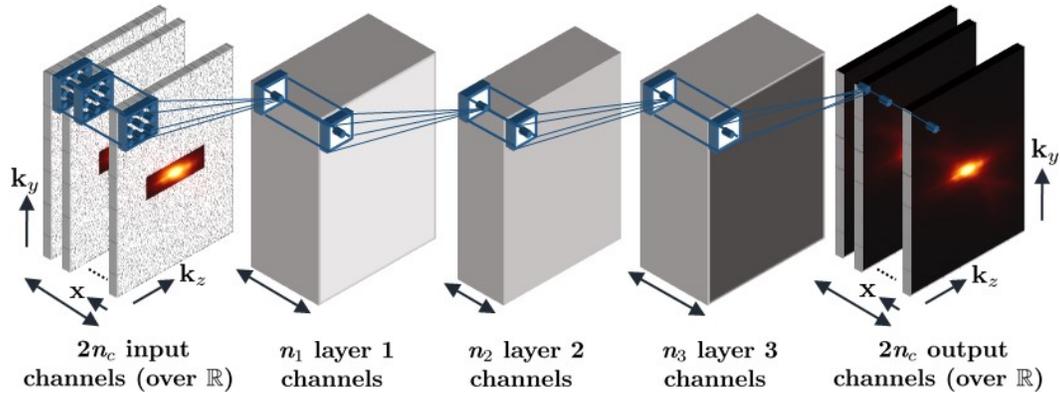

**Figure 1:** The CNN architecture to learn and enforce the coil self-consistency rule. The number of layer output channels is denoted by depth of blocks. All layers, except the last one, were followed by rectifier linear units (ReLU) as activation functions. The kernel sizes of the layers were 5×5, 3×3, 3×3 and 5×5, respectively. Each layer had 16, 8, 16 and $2n_c$ output channels, respectively.

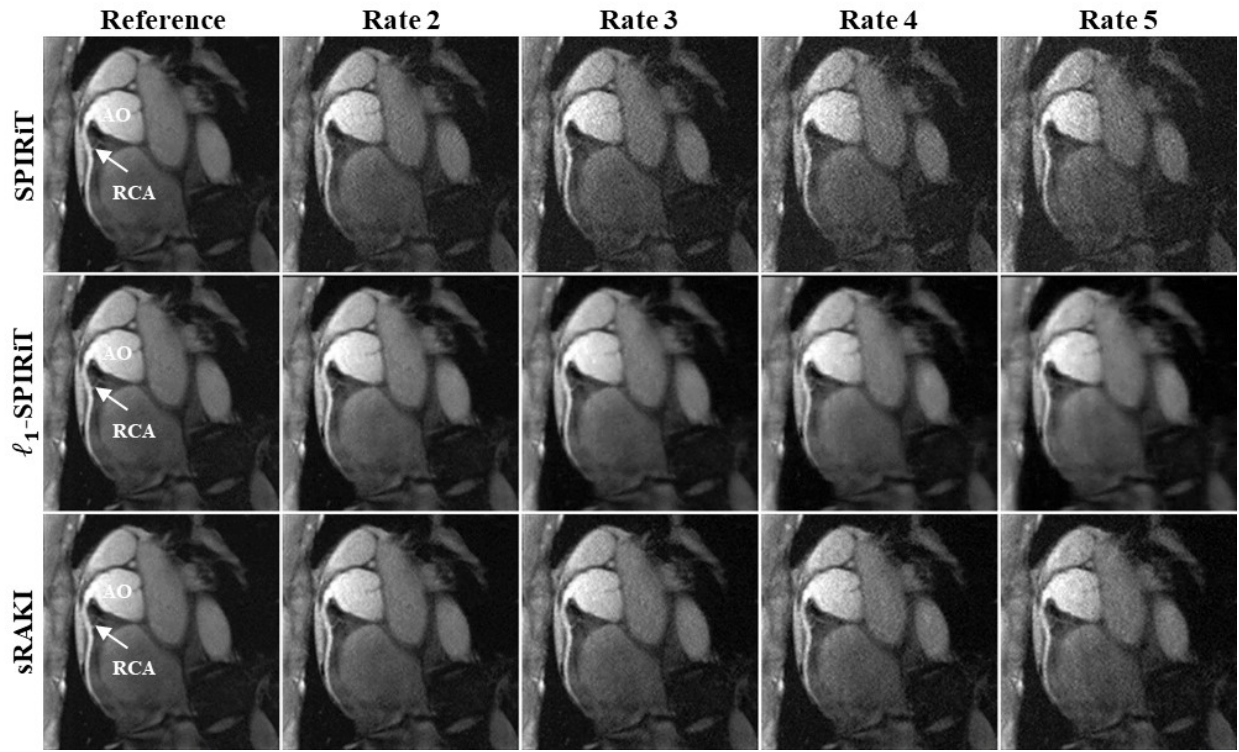

**Figure 2:** Reformatted right coronary artery (RCA) images from a 3D targeted coronary MRI dataset. The data were retrospectively undersampled at rates 2, 3, 4, and 5 in the $k_y - k_z$ plane and then reconstructed using SPIRiT, $\ell_1$-SPIRiT and sRAKI (top, middle and bottom rows). Fully-sampled images are also displayed in the first column as a reference for comparison. sRAKI is visually more robust to noise amplification and blurring artifacts at high acceleration rates compared to SPIRiT and $\ell_1$-SPIRiT, respectively. (RCA: right coronary artery; AO: Aortic Root)

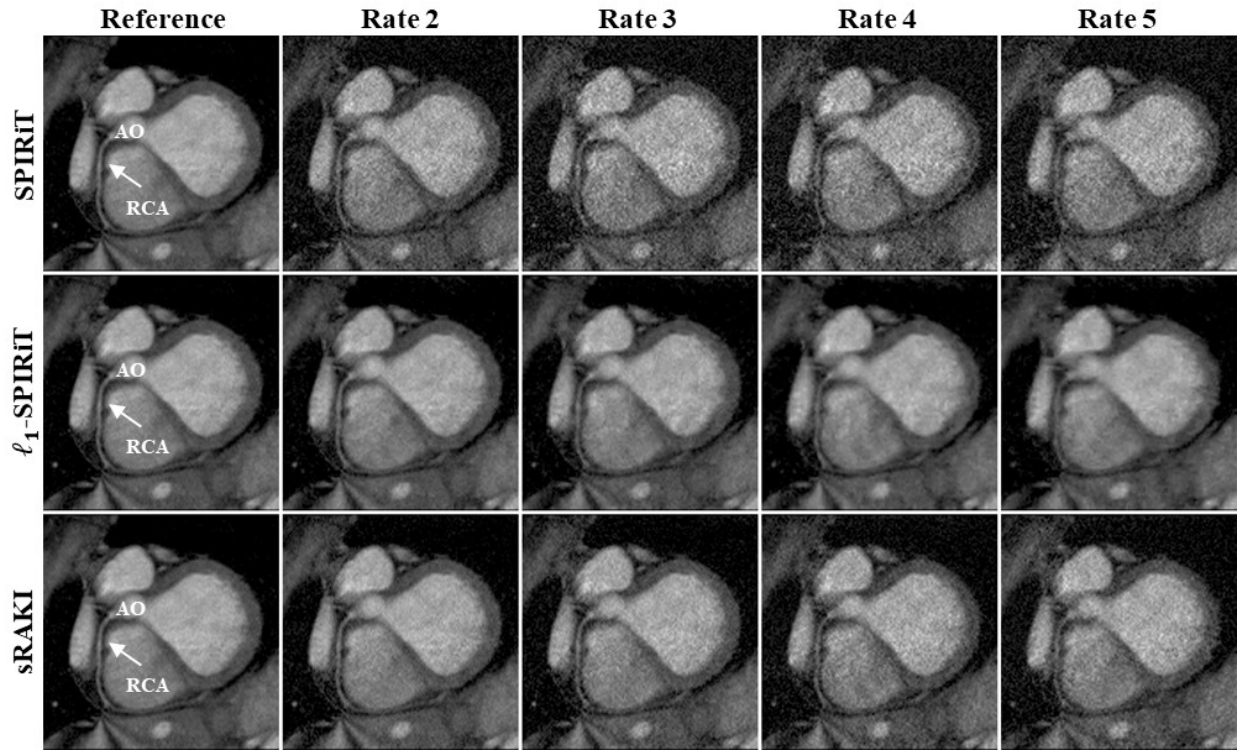

**Figure 3:** Reformatted right coronary artery (RCA) images from another 3D targeted coronary MRI dataset. This data were also retrospectively undersampled at rates 2, 3, 4, and 5, and fully-sampled images are shown in the first column as reference. The difference between SPIRiT and sRAKI is visually evident at all acceleration rates for this subject with more apparent noise amplification. Furthermore, compared to $\ell_1$-SPIRiT, sRAKI is more robust to blurring artifacts with increasing acceleration rates. (RCA: right coronary artery; AO: Aortic Root)

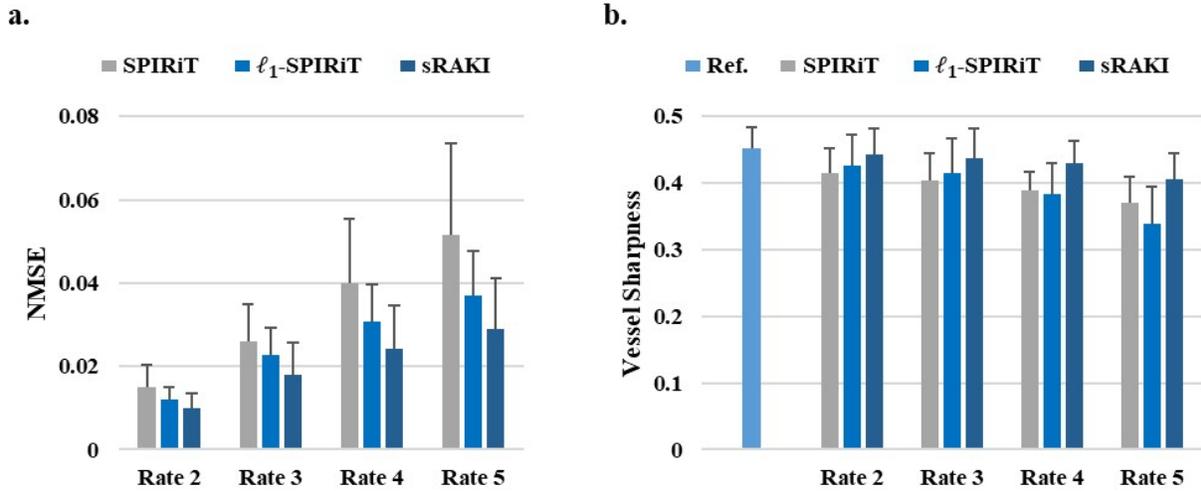

**Figure 4:** (a) Mean normalized mean squared error (NMSE) and (b) quantitative normalized vessel sharpness measures for all reconstructions of rates 2 to 5. Error bars represent standard deviation across subjects. sRAKI outperforms SPIRiT and $\ell_1$-SPIRiT at all rates for both metrics. The improvements in NMSE are statistically significant at all rates over SPIRiT, and at rates 4 and 5 over $\ell_1$-SPIRiT, whereas the improvements in vessel sharpness with sRAKI are significant at rates 2 and 4 over SPIRiT, and rates 4 and 5 over $\ell_1$-SPIRiT.

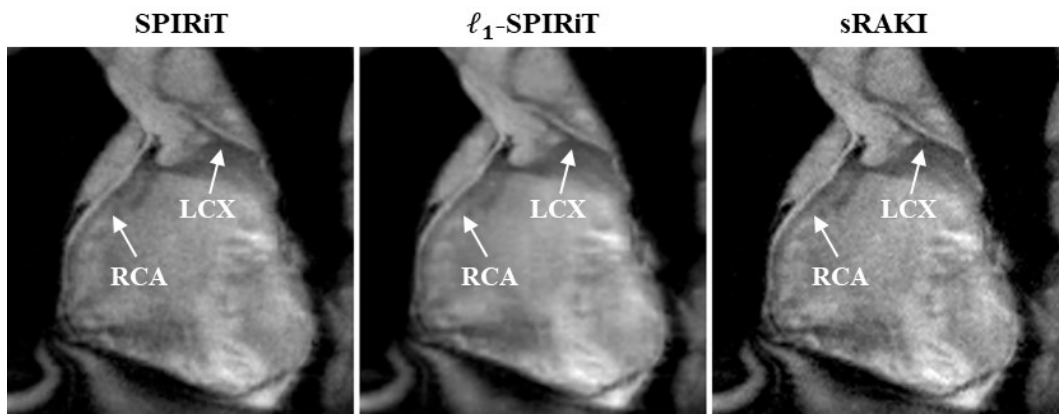

**Figure 5:** Reformatted coronal image from a prospectively 5-fold accelerated whole-heart coronary MRI dataset. The visualization of both the right coronary artery (RCA) and the left circumflex artery (LCX) is improved using sRAKI compared to SPIRiT and $\ell_1$-SPIRiT, with sharper definition of the arteries.